# Indirect Channel Sensing for Cognitive Amplify-and-Forward Relay Networks

Yipeng Liu and Qun Wan

*Abstract*—In cognitive radio network the primary channel information is beneficial. But it can not be obtained by direct channel estimation in cognitive system as pervious methods. And only one possible way is the primary receiver broadcasts the primary channel information to the cognitive users, but it would require the modification of the primary receiver and additional precious spectrum resource. Cooperative communication is also a promising technique. And this paper introduces an indirect channel sensing method for the primary channel in cognitive amplify-and-forward (AF) relay network. As the signal retransmitted from the primary AF relay node includes channel effects, the cognitive radio can receive retransmitted signal from AF node, and then extract the channel information from them. Afterwards, Least squares channel estimation and sparse channel estimation can be used to address the dense and sparse multipath channels respectively. Numerical experiment demonstrates that the proposed indirect channel sensing method has an acceptable performance.

## I. Introduction

COGNITIVE radio can dynamically sense the radio environment and rapidly tune their transmitter parameters to efficiently utilize the available channel source. And cognitive amplify-and-forward (AF) relay network is a main relaying protocol of cooperation communication for cognitive radio network (CRN) [1][2]. It is a powerful method proposed to achieve full utilization of robust dynamic spectrum by enabling a number of cooperating relay nodes to amplify and retransmit the received information. And it can support seamless data service for cognitive users while causing zero interference to primary systems.

Propagation channel is essential in cognitive systems [3] - [6]. An emitted signal is reflected, diffracted, and scattered in the environment, which makes the received signal as a superposition of multiple attenuated, delayed, and phase- and/or frequency-shifted copies of the original signal. It is the channel that significantly affects the signal quality in receiver and the corresponding communication performance. Besides,

Manuscript received MM. DD, YEAR. This work was supported by a grant from the National Natural Science Foundation of China (No.60772146) and the National High Technology Research and Development Program of China (863 Program, No.2008AA12Z306).

Yipeng Liu and Qun Wan are with the Electronic Engineering Department, University of Electronic Science and Technology of China, Chengdu, 610054 China. (e-mail: dr.yipengliu@gmail.com; wanqun@uestc.edu.cn).

Channel sensing is a vitally important part of the multi-dimensional spectrum sensing for cognitive communications in the grey spaces of spectrum [6]. With the sensed channel information, emitted signal can be designed orthogonal to the existing primary signals in multi-dimensions. Thus spectrum efficiency would be greatly enhanced. And channel information can assist positioning system too [6][20].

There are there types of channel in the cognitive AF relay network: The sensing channel is the one between primary transmitter (or primary AF relay transmitter) and local cognitive radio; and the reporting channel is the one between cognitive radio and the fusion center; and the primary channel is the one between the primary transmitter (or primary AF relay transmitter) and primary receiver (or primary AF relay receiver). In cognitive AF relay network, the first two kinds can be sensed directly by local cognitive radio as previous methods [3][7][8][9]. But the primary channel sensing in cognitive radio has not been explored well in the literature.

Here we set up the indirect channel sensing model for cognitive AF relay networks. The primary transmitter training signal is retransmitted from AF node, and then received in cognitive radio. As the signal received in AF node is infected by the primary channel, which means it contains the channel information, with the help of sensing channel estimation between the relay node and the cognitive radio, the transmitted signal in relay node can be estimated. Least squares (LS) estimation and sparse channel estimation are used to estimate the dense and sparse multipath channel impulse response respectively from the primary transmitted training signal and the retransmitted signal.

In the rest of this paper, Section II describes the primary channel model in cognitive AF relay networks. Section III provides the indirect channel sensing by LS estimation and sparse channel sensing successively. Section VI presents the simulation to demonstrate the methods and their performances. Finally conclusion is drawn in Section V.

## II. Channel Model

To obtain the primary channel information for cognitive radio in AF relay network, the cognitive radio can not perform the channel estimation directly as previous. As Fig. 1 shows, the primary channels CH1, CH4 can be sensed directly by primary AF relay node and receiver respectively. And the sensing channels CH2, CH3 and CH5 can be sensed directly by local cognitive radio. The channel information of the primary channel is also important to the local cognitive radio [3] - [6]. For example, primary channel information is able to enable knowledge-assisted positioning for cognitive users [20]. That is to say, the nodes' positions can be estimated by comparing the estimated channel parameters with the database. And it can help the agile signal design of cognitive radio according to the primary user's communication performance. But it is usually that the primary AF relay node or receiver can not broadcast the channel information of CH1 and CH4 to the local cognitive radio on another assigned channel, for it would incur additional channel resource burden and require the modification of primary receiver to broadcast. Thus the local cognitive radio should get the channel information indirectly by itself.

In the cognitive AF relay network, CH1 is estimated indirectly with the help of CH2 and CH3, and CH4 is done indirectly with the help of CH3 and CH5. As the CH1's sensing is the same with CH4's, here we only address the CH1.

Without loss of generality, assume the training sequence emitted by the primary transmitter is x(m), m = 0, ... , M−1 and the channel impulse response for CH1, CH2, CH3 are $h_1(n)$, $h_2(n)$, $h_3(n)$, n = 0, 1, ... , N −1, respectively. After the channel effects, the received signal samples in AF relay node can be modeled as an (M−1)×1 vector [8]:

$$\mathbf{y} = \mathbf{X}\mathbf{h}_1 + \mathbf{n}_1, \quad (1)$$

where

$$\mathbf{y} = [y(0), y(1), \cdots, y(M-1)]^T, \quad (2)$$

$$\mathbf{X} = \begin{bmatrix} x(0) & x(-1) & \cdots & x(-N+1) \\ x(1) & x(0) & \cdots & x(-N+2) \\ \vdots & \vdots & \ddots & \vdots \\ x(M-1) & x(M-2) & \cdots & x(-N+M) \end{bmatrix}, \quad (3)$$

$$\mathbf{n}_1 = [n_1(0), n_1(1), \cdots, n_1(M-1)]^T, \quad (4)$$

$$\mathbf{h}_i = [h_i(0), h_i(1), \cdots, h_i(N-1)]^T, i=1,2,3, \quad (5)$$

**X** is the M×N Toeplitz (convolutional) matrix, **y** is the (M−1)×1 observed vector in AF relay node, , $\mathbf{h}_i$ (i = 1, 2, 3) is the N×1 channel impulse response vector. $\mathbf{n}_i$ (i = 1, 2, 3) is the (M−1)×1 additive white Gaussian noise (AWGN) vector.

Similar to the primary signal samples in AF relay node (1), we model the sensing signal sample vectors **z** by way of CH3 as:
and **h** is the N×1 code rate selecting vector with K ones:

$$\mathbf{z} = \mathbf{H}_3 \mathbf{y} + \mathbf{n}_3, \quad (6)$$

where

$$\mathbf{z} = [z(0), z(1), \cdots, z(N+M-1)]^T, \quad (7)$$

$$\mathbf{n}_3 = [n_3(0), n_3(1), \cdots, n_3(N+M-1)]^T, \quad (8)$$

$$\mathbf{H}_3 = \begin{bmatrix} h_3(0) & 0 & \cdots & 0 \\ h_3(1) & h_3(0) & \cdots & 0 \\ \vdots & \vdots & \ddots & \vdots \\ h_3(M-1) & h_3(M-2) & \cdots & h_3(0) \\ \vdots & \vdots & \ddots & \vdots \\ h_3(N-1) & h_3(N-2) & \cdots & h_3(N-M) \\ 0 & h_3(N-1) & \cdots & h_3(N-M+1) \\ \vdots & \vdots & \ddots & \vdots \\ 0 & 0 & 0 & h_3(N-1) \end{bmatrix}, \quad (9)$$

**z** is the (M+N−1)×1 observed vectors in cognitive radio, $\mathbf{H}_3$ is an (M+N−1)×M Toeplitz (convolutional) matrix. $\mathbf{H}_3$ can be obtained in advance. The cognitive radio can transmit training signal to the relay node and receive the signal from the relay node, thus the channel information of $\mathbf{H}_3$ can be estimated.

## III. CHANNEL SENSING

It is often difficult for the CR to have a direct measurement of the primary channel [1]. Therefore, traditional spectrum sensing detect the primary signal based on the local observations of the CR. However, in the cognitive AF relay network, the cognitive radio can receive signals from the primary transmitter and the AF relay node on the sensing channel. The signal from the relay node carries the primary channel information. And we can obtain it indirectly from the local observations on the sensing channels.

### A. Indirect least Squares sensing

As it modeled in Section II, we formulate a set of equations:

$$\begin{aligned} \mathbf{y} &= \mathbf{X}\mathbf{h}_1 + \mathbf{n}_1 \\ \mathbf{z} &= \mathbf{H}_3\mathbf{y} + \mathbf{n}_3 \end{aligned}, \quad (10)$$

The primary channel sensing can boil down to solve $\mathbf{h}_1$ from (11). Assume the sensing channel information $\mathbf{H}_3$ is obtained directly in advance as section II explained. When $\mathbf{h}_1$ is of dense multipath as Fig. 2, LS channel estimation has a good performance. Under the assumption that $\mathbf{X}^H\mathbf{X}$ and $\mathbf{H}_3^H\mathbf{H}_3$ are invertible, these are simple linear estimators in the observations, where the LS estimators are given by

$$\mathbf{h}_1^{LS} = (\mathbf{X}^H\mathbf{X})^{-1}\mathbf{X}^H\mathbf{y}, \quad (11)$$

$$\mathbf{y}^{LS} = (\mathbf{H}_3^H\mathbf{H}_3)^{-1}\mathbf{H}_3^H\mathbf{z}, \quad (12)$$

Take (12) into (11), and the training signal is known in advance, the LS primary channel estimator is obtained.

## B. Indirect sparse channel sensing

LS channel estimator is the optimal for rich multipath channels. But recently numerous experimental studies have shown that practical wireless channels tend to exhibit sparse structures in the sense that majority of the elements in the channel impulse response vector are either zero or nearly zero when operating at large bandwidths and symbol durations and/or with large plurality of antennas [11]–[14].

First we obtain the primary transmitted signal and the retransmitted signal from AF node by LS estimation as before. Then we explore the sparsity of the primary channel CH1 and obtain the sparse indirect sensing.

$$\mathbf{h}_1^S = \arg\min_{\mathbf{h}_1}\left(\|\mathbf{y} - \mathbf{X}\mathbf{h}_1\|_2^2 + \lambda \|\mathbf{h}_1\|_p^p\right), \quad (13)$$

where $\lambda$ is a weighting factor controlling the tradeoff between $\|\mathbf{y} - \mathbf{X}\mathbf{h}_1\|_2^2$ and the degree of sparsity, which is represented by $\|\mathbf{h}_1\|_p^p$. $\|\mathbf{x}\|_p = \left(\sum_i |x_i|^p\right)^{1/p}$ is the $L^p$ norm of $\mathbf{x}$, The smaller the $p$, the sparser the selecting vector. When $0 \leq p \leq 1$, the $L^p$ norm can be described as a diversity measure [18]. The smaller the value of $p$ is, the sparser the selecting vector $\mathbf{h}_1$ is, meaning that the number of trivial entries in $\mathbf{h}_1$ is larger.

To obtain the optimal $\mathbf{h}_1$, we define the cost function as

$$J(\mathbf{h}_1) = \|\mathbf{X} - \mathbf{A}\mathbf{h}_1\|_2^2 + \lambda \|\mathbf{h}_1\|_p^p, \quad (14)$$

The gradient of $J(\mathbf{h}_1)$ with respect to $\mathbf{h}_1$ is given by

$$\nabla_{\mathbf{h}} J(\mathbf{h}_1) = 2\mathbf{A}^H \mathbf{A}\mathbf{h}_1 - 2\mathbf{A}^H \mathbf{X} + \lambda' \Pi(\mathbf{h}_1)\mathbf{h}_1, \quad (15)$$

where

$$\lambda' = \lambda p, \quad (16)$$

$$\Pi(\mathbf{h}_1) = diag\left\{|h_1[1]|^{p-2}, \cdots, |h_1[i]|^{p-2}, \cdots |h_1[N]|^{p-2}\right\}, \quad (17)$$

and $h_1[i]$ denotes the $i$th element of $\mathbf{h}_1$.

Setting (15) to zero and following similar approaches as in [18], the recursive updating formula for $\mathbf{h}_1$ is given by

$$\mathbf{h}_1^{k+1} = \mathbf{W}_{k+1} \mathbf{A}_{k+1}^T \left(\mathbf{A}_{k+1} \mathbf{A}_{k+1}^T + \lambda' \mathbf{I}\right)^{-1} \mathbf{X}, \quad (18)$$

where $\mathbf{h}_1^k$ is the selecting vector at the $k$th step.

$$\mathbf{W}_{k+1} = diag\left\{|h_1^k[1]|^{(1-p/2)}, \cdots, |h_1^k[N]|^{(1-p/2)}\right\}, \quad (19)$$

$$\mathbf{A}_{k+1} = \mathbf{A}\mathbf{W}_{k+1}, \quad (20)$$

In (13), $p$ and $\lambda$ determine the effectiveness of the sparsity constraint on the selecting vector $\mathbf{h}_1$. For a fixed $\lambda$, the smaller the value of $p$ is, the sparser the obtained selecting vector $\mathbf{h}_1$ is. When $p$ is fixed, principles of choosing a proper $\lambda$ have been proposed in [18].

In order to reduce the computational intensity, we can assume $p = 1$ in (13) for computation convenience, for if and only if $p = 1$, (13) is a convex optimization problem. Then for which the optimal value of $\mathbf{h}_1$ can be obtained readily using available software such as CVX [19]. As sparse channel sensing explores the sparsity, which means it uses more information than a traditional LS estimator does, it is expected to achieve better estimation performance.

## IV. SIMULATION

The numerical experiment is used to demonstrate the performance of the proposed method. All the channel impulse response is set to 100 long, and the training data is a serial of 50 consecutive numbers. The sensing channels are generated with the exponential fading as Fig 2. The primary channel is assumed to be a sparse multipath channel with the number of nonzero taps is 15. And most of the nozero taps are located in the first 10 positions in the channel model. And Nonzero taps are randomly distributed in [−1, −0.2] ∪ [0.2, 1]. An example of the sparse channel is Fig 3.

1000 Monte Carlo simulations with different signals and the nonzero taps are performed by cumulative density function (CDF). Root mean square error (RMSE), which uses these estimation methods, is defined as

$$RMSE = \sqrt{\frac{1}{L}\|\hat{\mathbf{h}}_1 - \mathbf{h}_1\|_2^2}, \quad (21)$$

where $\mathbf{h}_1$ and $\hat{\mathbf{h}}_1$ are the real and estimated channel impulse response vectors respectively, L is the number of Monte Carlo Simulations. Here we let L = 1000.

Fig 4 and Fig. 6 illustrate the overall taps estimation RMSE and the nozero taps estimation RMSE vs cumulative density function (CDF) by LS and sparse channel estimation respectively, when the signal to noise ratios (SNRs) over all the channels CH1, CH2 and CH3 are set to be SNR1 = SNR2 = SNR3 = 10dB. It can be seen that the primary channel sensing performance on the assumed SNR condition. As the primary channel is assumed to be sparse, the sparse channel estimation outperforms the LS channel estimation. And the indirect sparse channel sensing result can be acceptable. Besides, the RMSE performance of nonzero tap estimation is inferior to overall tap estimation's performance. And the sparse channel estimation of nonzero tap is superior to the LS estimation.

Fig.5 and Fig. 7 demonstrate the performance when SNR1 = SNR2 = SNR3 = 20dB. As the SNR increases, both the RMSE performances of overall tap estimation and nonzero tap estimation improve significantly. And the sparse channel estimation's superiority to LS channel estimation for the sparse channel is further enhanced.

## V. Conclusion

Channel estimation is a hot research topic in signal processing for wireless communication. But the indirect primary channel sensing in the cognitive AF network is not discussed in the literature. Here the authors set up the transceiver signal model in the cognitive AF network, and propose a method to do the primary channel sensing indirectly by the cognitive engine merely. The LS method and sparse method are incorporated to do the estimation. It does not require either additional spectrum resource or some modification on primary transceiver network. Simulation demonstrates the methods and the corresponding performance is acceptable.

Here we only take two main channel estimation methods as examples. And channel estimation can be obtained by other sparse channel estimations, such as Partial sparse multi-path channel estimation [9], compressed channel sensing [15][17], etc. Furthermore, usually there is not only one cognitive user in the network, and the information is shared in the cognitive network. Thus, the primary channel can be estimated by more than one node, and the performance gain can be gotten by fusion of the results of cognitive radios.

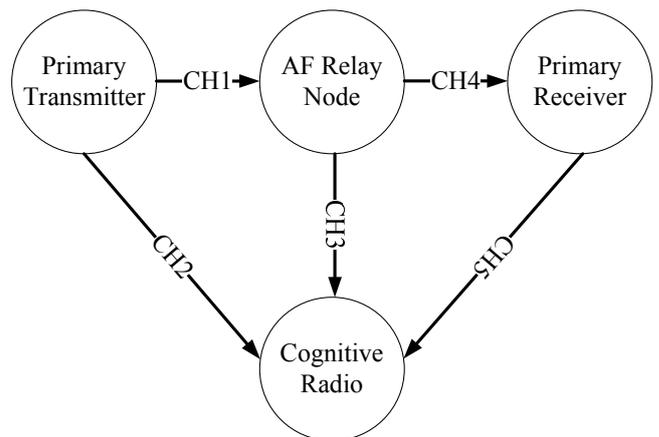

Fig. 1 A brief model of cognitive AF relay network.

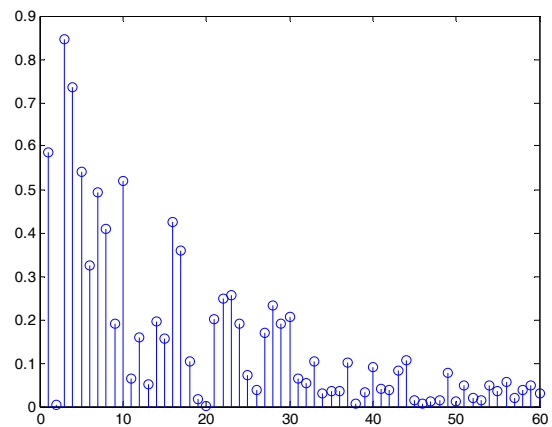

Fig. 2    A sample of dense multipath channel.

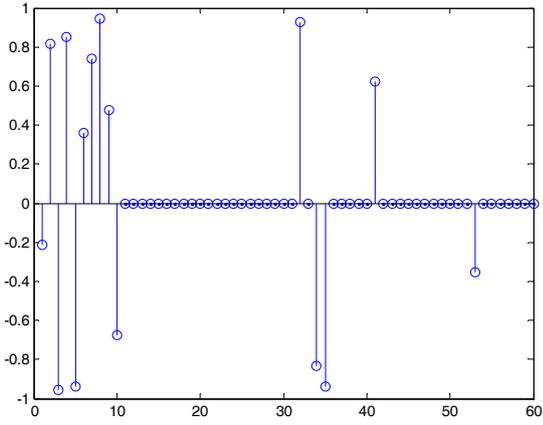

Fig. 3  A sample of sparse multipath channel.

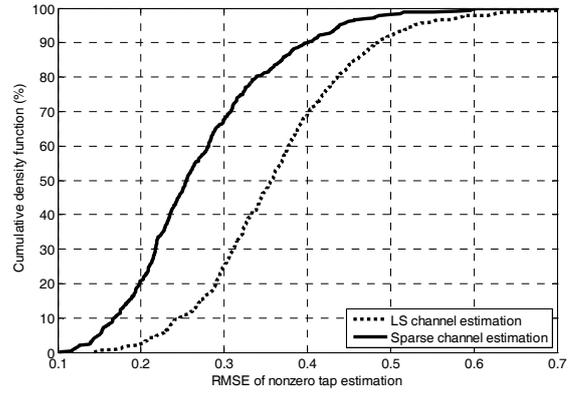

Fig. 6 RMSE of nonzero tap estimation SNR1= SNR2= SNR3=10dB

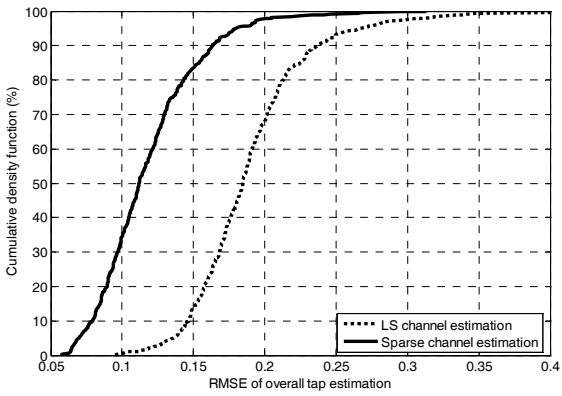

Fig. 4 RMSE of overall tap estimation SNR1= SNR2= SNR3=10dB

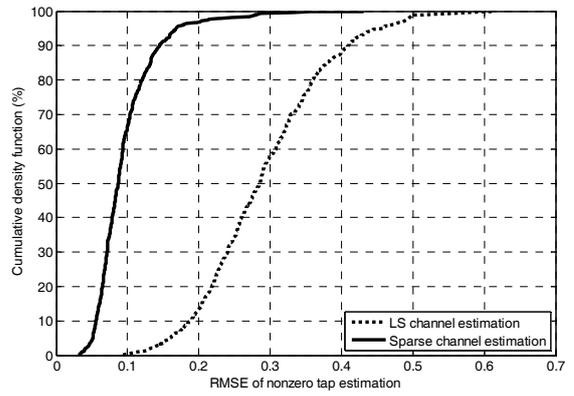

Fig. 7 RMSE of nonzero tap estimation SNR1= SNR2= SNR3=20dB

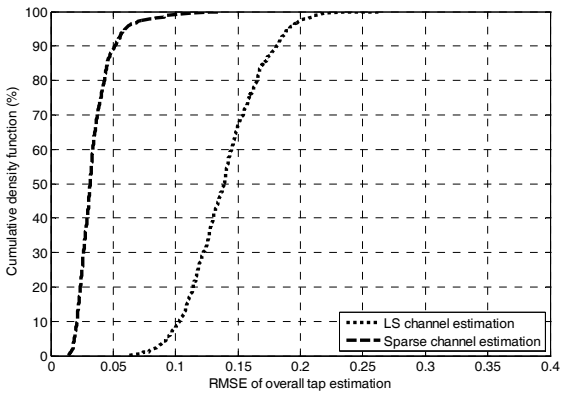

Fig. 5 RMSE of overall tap estimation SNR1= SNR2= SNR3=20dB